\documentclass[twoside,a4paper,11pt]{proceedings}
\usepackage{graphicx}
\usepackage{hyperref}
\usepackage{movie15}
\usepackage{natbib}
\usepackage{color}  

\begin{document}
\pagenumbering{arabic}
\pagestyle{myheadings}
\thispagestyle{empty}
\vspace*{-1cm}
{\flushleft\includegraphics[width=8cm,viewport=0 -30 200 -20]{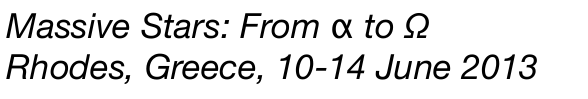}}
\vspace*{0.2cm}
\begin{flushleft}
{\bf {\LARGE
On the nature of the WO3 star DR1 in IC\,1613
}\\
\vspace*{1cm}
F. Tramper$^1$,
G. Gr\"afener$^2$,
O.E. Hartoog$^1$,
H. Sana$^3$,
A. de Koter$^{1,4}$,
J.S. Vink$^2$,
L.E. Ellerbroek$^1$,
N. Langer$^5$,
M. Garcia$^6$,
L. Kaper$^1$
and
S.E. de Mink$^{7,8}$
%
}\\
\vspace*{0.5cm}
%
$^{1}$
Astronomical Insitute 'Anton Pannekoek', University of Amsterdam, The Netherlands \\
$^{2}$
Armagh Observatory, Northern Ireland \\
$^{3}$
Space Telescope Science Institute, Baltimore, USA \\
$^{4}$
Insituut voor Sterrenkunde, Universiteit Leuven, Belgium \\
$^{5}$
Argelander Institut f\"ur Astronomie, University of Bonn, Germany \\
$^{6}$
Centro de Astrobiologica, CSIC-INTA, Madrid, Spain \\
$^{7}$
Observatories of the Carnegie Institution for Science, Pasadena, USA\\
$^{8}$
Cahill Center for Astrophysics, California Institute for Technology, Pasadena, USA \\

%
\end{flushleft}
\markboth{
DR1
}{
Tramper et al.
}
\thispagestyle{empty}
\vspace*{0.4cm}
\begin{minipage}[l]{0.09\textwidth}
\ 
\end{minipage}
\begin{minipage}[r]{0.9\textwidth}
\vspace{1cm}
\section*{Abstract}{\small
We present the results of a quantitative spectroscopic analysis of the oxygen-sequence Wolf-Rayet star DR1
in the low-metallicity galaxy IC\,1613.  Our models suggest that the strong oxygen emission lines are the result
of the high temperature of this WO3 star and do not necessarily reflect a more advanced evolutionary stage 
than WC stars.
\vspace{10mm}
\normalsize}
\end{minipage}

\section{Introduction}

Oxygen-sequence Wolf-Rayet (WO) stars are extremely rare: only 8 are currently known. They are often thought to represent the evolutionary stage succeeding the carbon-sequence Wolf-Rayet (WC) phase \citep{smith1991}. The strong oxygen emission then originates from the surfacing of this species at the end of core-helium burning. An alternative explanation is that the high excitation oxygen emission reflects a higher stellar temperature compared to WC stars \citep{crowther1998}. Here, we attempt to discriminate between these scenarios by analyzing
the optical to near-inrared spectrum of the WO3 star DR1 in IC~1613. Our full analysis is provided in \cite{tramper2013}.

\section{Modeling DR1 in IC\,1613}

DR1 is a WO3 star located in the low-metallicity Local Group galaxy IC~1613. The metallicity of this system is
$Z \sim 1/7 Z_{\odot}$ \citep[][]{bresolin2007}, making DR1 the only WO star known in a sub-Small Magellanic Cloud
metallicity environment (with $Z_{\rm SMC} = 1/5 \, Z_{\odot}$). 
Using {\sc CMFGEN}, we model the observed spectrum of DR1 to determine its stellar and outflow properties, in particular the oxygen and carbon abundance and the stellar temperature.

\begin{figure}
\center
\includegraphics[width=0.90\textwidth]{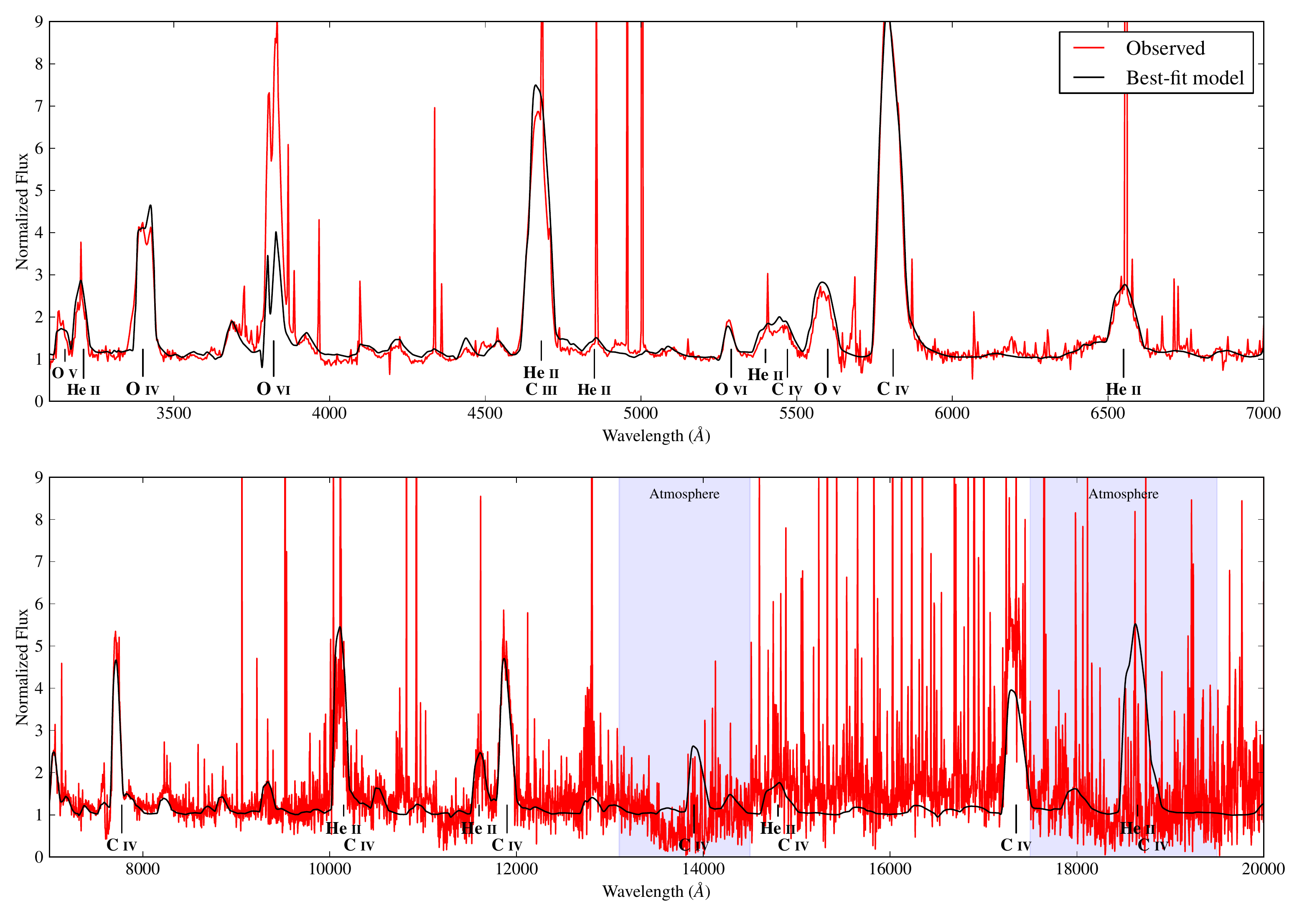}
\caption{VLT/X-Shooter spectrum of DR1 in red. Overplotted in black is the best-fitting model. The shaded areas indicate wavelength regions where the Earth's atmosphere is opaque.}\label{fig:bestmod}
\end{figure}
\begin{figure}
\center
\includegraphics[width=0.69\textwidth]{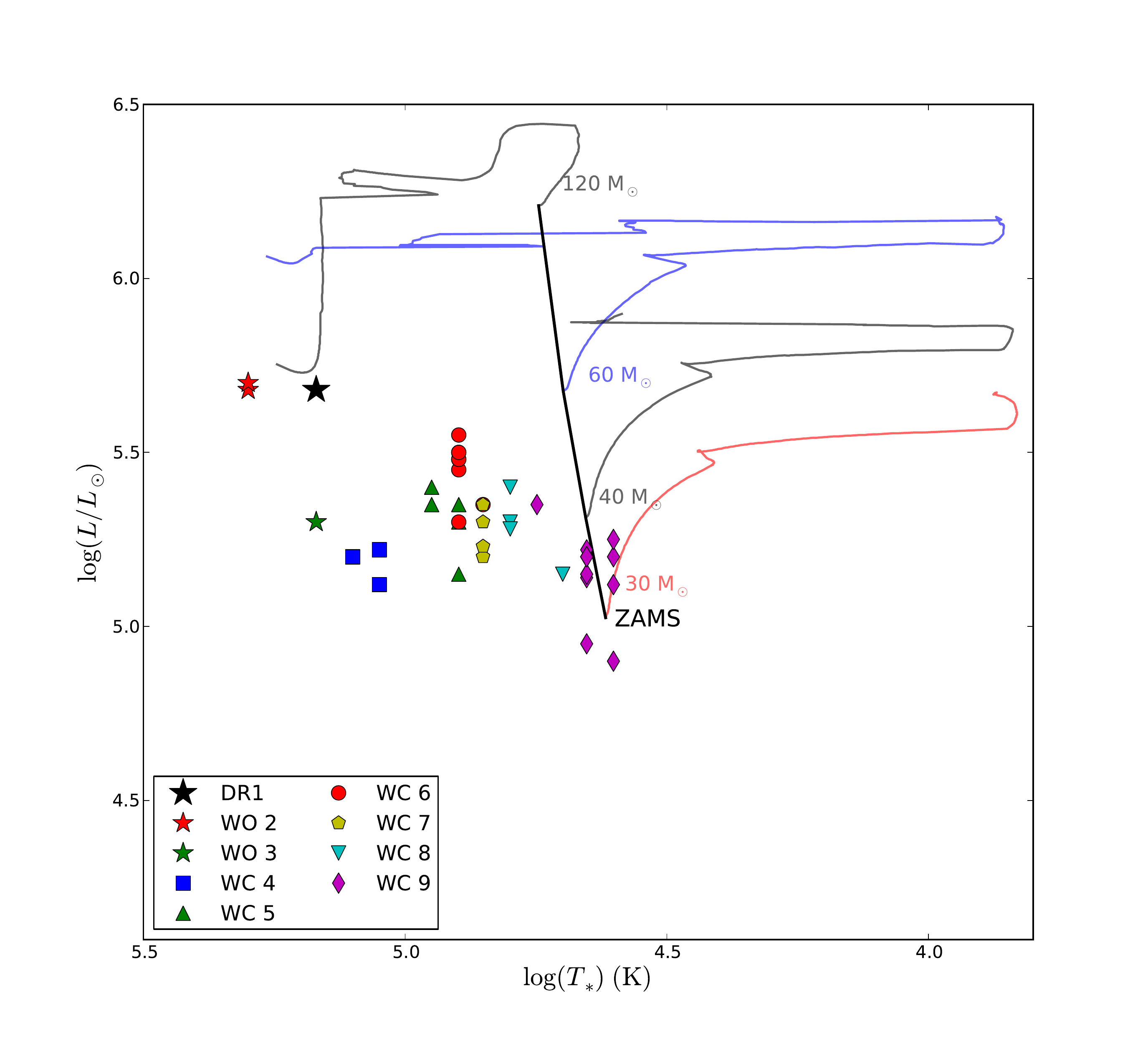}
\caption{Hertzsprung-Russell diagram indicating the location of DR1 and several other WO and WC stars analyzed by \cite{sander2012}. Also plotted are evolutionary tracks from \cite{maeder2005} for a metallicity
characteristic of the Small Magellanic Cloud.}\label{fig:HRD}
\end{figure}

Our best model represents the spectrum of DR1 well safe for the O{\sc vi} $\lambda\lambda$3811-34 emission, the strength of which is not fully reproduced (see Figure~\ref{fig:bestmod}). DR1 has a luminosity that is $10^{5.74 \pm 0.10}$ times that of the sun and a temperature of about $150 \pm 25$\,kK. Both these values are high
compared to WC stars (see Figure~\ref{fig:HRD}). We suggest that the poorly fitted O{\sc vi} $\lambda\lambda$3811-34 emission originates from a mechanism not accounted for in our model (possibly shock induced X-rays). Models that can reproduce this line need a high temperature and high oxygen abundance, but do not fit the rest of the spectrum.

\begin{figure}
\center
\includegraphics[width=0.60\textwidth]{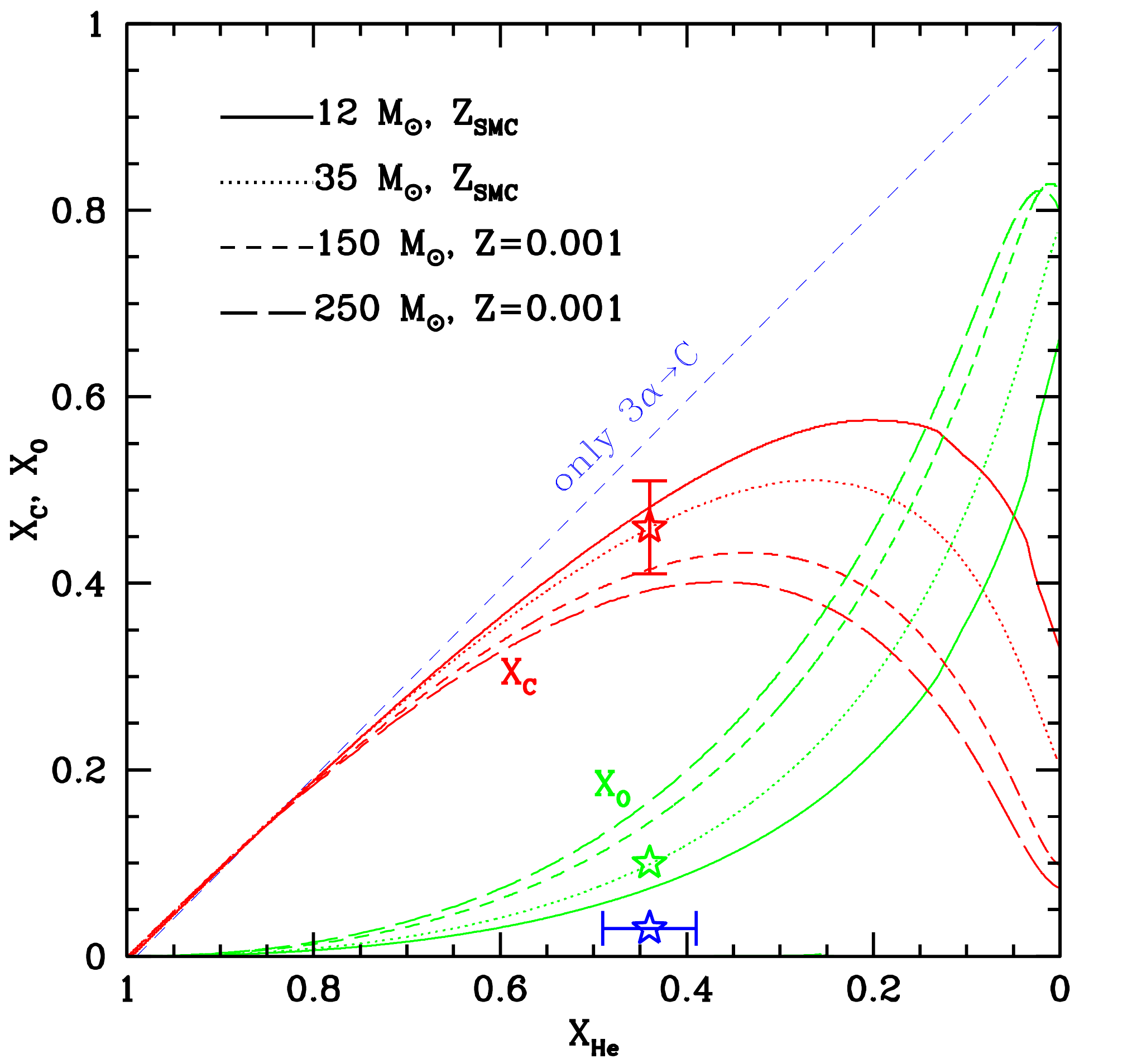}
\caption{Surface abundances predicted for helium-burning stars \citep{langer2007, brott2011}. The star symbols indicate the surface abundances of carbon (in red), oxygen (in green) and helium (in blue) found for DR1.}\label{fig:abundance}
\end{figure}

We derive an oxygen abundance relative to helium of O/He = $0.06 \pm 0.01$ by number and a carbon abundance C/He = $0.45 \pm 0.05$. These values are comparable to those found for WC stars in various environments.  The 
current surface abundances, including the mass fraction of helium, are plotted in Figure~\ref{fig:abundance}. Our abundances suggest that DR1 is more than half-way through its core-helium burning stage \citep[see][for details]{tramper2013}.

\section{Discussion and conclusions}

The oxygen abundance is not enhanced compared to values found for WC stars in other studies. The temperature and luminosity, however, are higher. If DR1 is representative for the WO class, this would suggest that WO stars are not necessarily in a more advanced evolutionary stage than WC stars.  Figure~\ref{fig:HRD} shows that DR1 is located at a position
in the HRD expected for the late stages of evolution of a 120 $M_{\odot}$ star at SMC metallicity, in which case its
current mass would be about 18 $M_{\odot}$.  Stars with a final mass larger than about 10 solar masses are likely
to form black holes, producing a faint supernova or no supernova at all.  If rapidly rotating, however, they may produce
bright type Ib/c supernovae.  Although DR1 still contains an appreciable amount of helium -- 44\% at the surface -- this does not exclude that DR1 may end in a type Ic supernova.  Supernova progenitors may have a surface helium mass fraction as large as 50 percent without this species being detected in the spectrum of the supernova \citep{dessart2011}.

\small  
%

\bibliographystyle{aj}
\small
\bibliography{proceedings}

\end{document}